# Discovery of Niobium Hydride Precipitates in Superconducting Qubits

Zuhawn Sung,[1*†] Daniel Bafia,[1*] Arely Cano,[1] Akshay Murthy,[1] Jaeyel Lee,[1] Matthew J Reagor,[2] Juan Rubio-Zuazo,[3,4] Anna Grassellino,[1] and Alexander Romanenko[1]

[1]Fermi National Accelerator Laboratory, Batavia, IL, 60510, USA
[2]Rigetti Computing, Berkeley, CA, 94710, USA
[3]Bm25-SpLine at European Synchrotron Radiation Facility, 38043 Grenoble, France
[4]Instituto de Ciencia de Materiales de Madrid – CSIC, Cantoblando, Madrid, Spain

**ABSTRACT**. We report the evidence of the formation of niobium hydride phase within niobium films on silicon substrates in superconducting qubits fabricated at Rigetti Computing. For this study, we combined complementary techniques—including room-temperature and cryogenic atomic force microscopy (AFM), synchrotron X-ray diffraction, and time-of-flight secondary ion mass spectroscopy (ToF-SIMS)—to directly reveal the existence of niobium hydride precipitates in the Rigetti chip area. Upon cryogenic cooling, we observed variation in the size and morphology of the hydrides, ranging from small (~5 nm) irregular shapes to large (~10-100 nm) domain within the Nb grains, fully converted to niobium hydrides. Since niobium hydrides are non-superconducting and can easily change in size and location upon different cooldowns to cryogenic temperature, our finding highlights a new and previously unknown source of decoherence in superconducting qubits. This contributes to both quasiparticle and two-level system (TLS) losses, offering a potential explanation for changes in qubit performance upon cooldowns. Finally, by leveraging the RF performance of a 3D bulk Nb resonator, we can quantify RF dissipation on a superconducting qubit, caused by hydrogen concentration variation, and are able to propose a practical engineering pathway to mitigate the formation of the Nb hydrides for superconducting qubit applications.

## I. INTRODUCTION.

Superconducting niobium has gained widespread recognition for transmon qubit applications in quantum computing. It presents the highest $T_c$ among pure metals as a gapped superconductor and shows anisotropies regarding material properties [1–3]. Niobium thin films, typically ~100-200 nm thick, are key components in many superconducting qubit architectures, used to form readout resonators, coupling lines, and capacitance pads present in qubits [4,5]. Therefore, any microwave dissipation mechanisms present in the niobium film itself, or native oxide layers on the top of the film, amorphous interface layers between the film and substrate, and the substrate are contributors to the decoherence of the superconducting qubits. There have been intensive studies to understand the microscopic origins of the decoherence in the superconducting qubits [4,6], and several coherence-limiting mechanisms have been identified [7]. Those are amorphous oxides, which host a two-level system (TLS) at various interfaces [4,8–10], metal-air [3,11,12], metal-substrate [13–15], and substrate-air interface [3,14], which have been recently shown to be the primary coherence-limiting mechanism in the quantum regime for 3D superconducting radio frequency (SRF) resonators (cavities) [12,16]. Additionally, non-equilibrium quasiparticles [17], including those produced by cosmic rays and radioactive sources in the environment [18,19], are another possible decoherence mechanism.

It has been previously discovered in bulk niobium SRF cavities that the formation of the non-superconducting niobium nano-hydrides upon cooldown from room temperature to cryogenic temperatures (< 150 K) introduces a major dissipation mechanism, which limits the achievable quality factor [20–23]. The process is enabled by Nb's ability to absorb significant amounts of hydrogen (H) even at room temperature, whenever a layer of native niobium oxide, mainly $Nb_2O_5$, does not passivate it [24,25]. Hydrogen atoms can occupy 12 tetrahedral sites in the body-centered cubic (bcc) Nb lattice, and they can induce several crystalline hydride phase formations linked to the occupied atomic position within the Nb lattice [26,27], particularly at temperatures below the saturation point of ~ 150 K [28]. The hydride phase can limit the achievable quality factors of the resonators [29] as significant microwave dissipation arises from the local suppression of superconductivity via proximity effects [22]. The possible presence and dissipation due to niobium hydrides, well-studied in bulk Nb SRF cavities, has not been extensively studied yet for the 2D superconducting qubits.

In this study, we apply the same techniques used to reveal the niobium hydrides in the bulk SRF Nb cavities to niobium film regions of the superconducting qubit test chips fabricated by Rigetti Computing [30]. We combine atomic force microscopy (AFM), synchrotron X-ray diffraction, and the time-of-flight secondary ion mass spectroscopy (ToF-SIMS) at room and cryogenic temperatures to reveal the niobium hydrides' existence directly in the Rigetti superconducting qubits. In parallel, the microwave performance of a bulk SRF resonator is examined at a low RF field regime (< 1 MV/m) and low temperature below 1.5 K after varying amounts of hydrogen loading content to assess the impact of Nb hydride on the extent of decoherence of superconducting qubits. These newly identified niobium hydrides may explain variability in qubit performance across cooldown cycles. Additionally, we discuss strategies to mitigate Nb hydride formation in superconducting qubit applications.

†zsung@fnal.gov

## II. RESULTS

Superconducting qubit test devices with Nb resonators and Al/AlOx/Al Josephson junctions were fabricated on a Si (100) substrate (float-zone >10,000 Ω cm) at Rigetti Computing following the published nanofabrication procedure [30]. The chip contains five superconducting qubits, readout resonators, and connecting lines. The Nb films on the test chips were deposited via HiPIMS at room temperature, with a base pressure below $1 \times 10^{-8}$ Torr.

In the inset of Fig. 1(a), a schematic of the superconducting qubit chip analyzed in this study is displayed with the Si substrate, Al/AlO$_x$/Al Josephson junction area, and Nb film pads labelled accordingly. The surface morphology of the Nb film is first analyzed using atomic force microscopy (AFM) at room and cryogenic temperatures, as shown in Fig. 1(a-d). AFM scan was performed at every 50 K step from 300 K down to 2 K with a cooling rate of 20 K/min, followed by a 3-hour extended stay to avoid thermal fluctuation. At room temperature (RT), the Nb surface is smooth, displaying root-mean-square (RMS) roughness values <1 nm. However, we observed two distinct morphological features that arise during cooling down to 2 K and warming up back to 300 K. Irregularly shaped structures of ~ 500 nm lateral diameter and 10-20 nm of height appeared on the surface of the Nb films as the temperature approaches 200 K after the initial cooling down to 2 K and persisted throughout warming up to 250 K as shown in Fig. 1(a, b). These features disappeared entirely at 300 K. Additional topological features with the size on the order of ~ 2 nm are observed to emerge in some of the Nb grains as well from initial cooldown below 50 K, Fig. 1 (c, d). In the case of bulk Nb SRF cavities, similar surface features have been previously detected at cryogenic temperatures and have been shown to result from the precipitation of the Nb hydride [31].

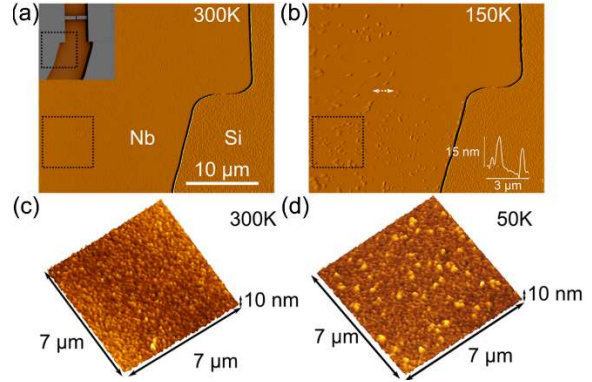

FIG 1. (a) Inset - schematic of the Nb qubit pads area, cryo-AFM area of analysis is shown by dashed rectangle; AFM surface images of Nb films and Nb/Si boundary at room temperature, and (b) at T = 200 K (during warming up back to 300 K from initial cooling down to 2 K), with the formation of the unknown structures on the surface of Nb clearly observed; the line profile through the structures at the white arrow is shown in the inset on the right, indicating the height of the order of 15 nm. The higher magnification surface topology of Nb films from the dashed rectangles in (a) and (b) is shown in (c) RT and (d) 50 K (during initial cooldown to 2 K), respectively, with further smaller scale formations apparent.

ToF-SIMS was used to probe the chemical identity associated with forming these surface features. Specifically, depth profile measurements performed at room temperature on several Nb contact pad regions from the same qubit device revealed an appreciable level of hydrogen at the surface [Fig. 2 (a)]. We find that the H-/Nb- signal, which represents a measure of the free hydrogen concentration present in the parent Nb matrix, is maximum immediately beneath the niobium oxide layer, Nb$_2$O$_5$, and then decays to a much lower level within the first ~10 nm depth from the surface. The presence of this hydrogen at the surface is consistent with previous findings in the bulk Nb cavities and also potentially drives Nb hydride formation in Nb thin film geometries. Specifically, compared to a bulk Nb cavity, the H-/Nb- signal within the top 25 nm is more prominent in the contact pad [Fig. 2(b)]. In addition to hydrogen, significant concentrations of other impurities are observed. For example, the O-/Nb- and C-/Nb- signals, which also decay on a longer length scale than the H-/Nb- signal, are shown in Fig. 2 (a).

†zsung@fnal.gov

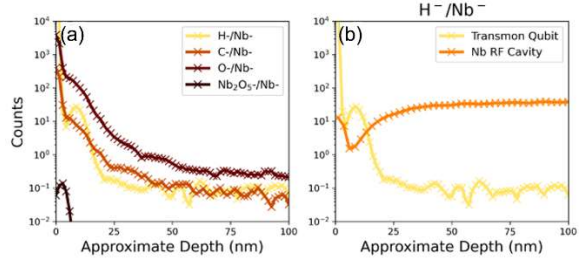

FIG 2. (a) SIMS depth profiles of O, H, C impurities and the $Nb_2O_5$ surface oxide layer in Nb film in the superconducting qubit. (b) The comparison of hydrogen impurity in the transmon qubit and the bulk Nb surfaces.

Cryo-grazing incidence X-ray diffraction (GIXRD) was performed at the Spanish beamline BM25 (SpLine) of the European Synchrotron Research Facilities (ESRF) to identify Nb hydride precipitates in the Nb film. This film was co-fabricated with the Regetti qubit device using HiPIMS. Diffraction patterns were collected at energies up to 18.34 keV to avoid the Nb K$\alpha$ fluorescence (20-25 keV) [32]. The Nb film was mounted on a Cu cold finger under ultra-high vacuum (UHV) conditions (~$10^{-9}$ mbar), allowing cooling down to 5 K via a continuous liquid helium (He) stream. The X-ray penetration depth ($\lambda = 0.68$ Å) was estimated using Parratt's equation [33] for an incident angle range of 0.15° to 2°. In Fig. 3, the dominant crystallographic reflection corresponds to Nb (101) plane at d = 2.40 Å. However, a new diffraction peak emerges in the 150 K to 100 K temperature range [Fig. 3(b)], which is absent at higher temperatures (300 K), as shown in Fig. 3(a). This reflection corresponds to the (111) plane of the orthorhombic Nb-hydride phase, $NbH_{0.89}$, in *Pnnn* space group, at d = 2.45 Å. This hydride-related peak shifted slightly by ~ 0.02Å upon a subsequent cooldown but maintained its intensity between 100 K and 150 K. Interestingly, no other reflections related to hydrides were observed upon further cooling to 5 K, and the hydride reflections completely disappeared after warming back to 300 K. This behavior is analogous to previous observations in the bulk SRF Nb cavities, suggesting that the surface topographical features observed by cryo-AFM indeed corresponds to hydride phases.

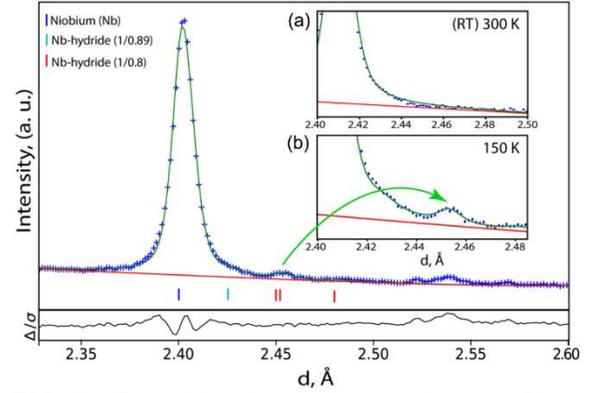

FIG 3. Rietveld refined the synchrotron X-ray diffraction pattern of a Nb film measured at 150 K. The insets show the interest region at (a) 300 K and (b) 150 K. At 150 K, we observed the presence of weak peaks associated with Nb-hydride precipitations.

Next, we employed a 3D 1.3 GHz superconducting radio-frequency (SRF) niobium resonator (cavity) as a simplified, single-interface system to quantify RF dissipation caused by varying hydrogen concentrations in the RF penetration layer. Initially, the baseline test of the 1.3 GHz Nb cavity was performed after following an empirical engineering process of the cavity surface treatments. The cavity was tested with power balance RF measurement to extract the unloaded quality factor, $Q_0$ [34], and applied to zero-span decay measurement as discussed in Ref [10,16] to probe the quality factor at the lower field regime. After the baseline testing, hydrogen incorporation was performed by mechanical grinding the cavity's exterior with a vibrational sander under continuous deionized water flow, maintaining the resonator's inner surface integrity. To further enhance hydrogen uptake and elevate hydrogen concentration within the cavity inner wall, an additional H-loading procedure was executed with electrochemical treatment. The details of cavity preparation are described in the supplementary materials. Fig. 4 summarizes the results of the RF tests. To verify successful hydrogen incorporation, two representative cavity cutouts underwent identical H-loading procedures, and their hydrogen concentrations were analyzed using ToF-SIMS. These results, also displayed in Figure 4, confirm increased hydrogen concentrations.

†zsung@fnal.gov

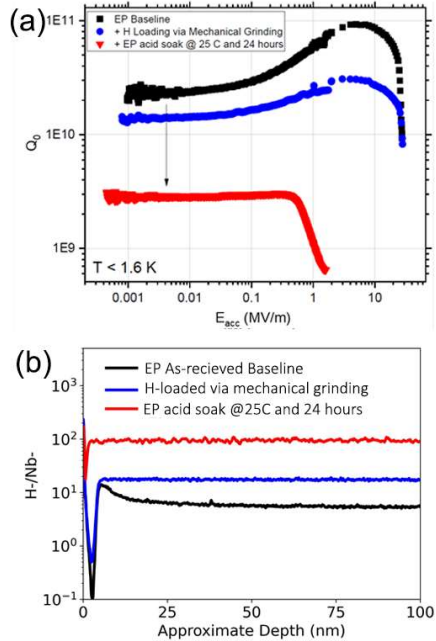

FIG 4. (a) RF measurements at a temperature less than 1.6 K on a 1.3 GHz niobium SRF resonator subjected to various hydrogen loading processing techniques. (b) Tof-SIMS acquired data on cavity cutouts of EP as-received, after H-loading via mechanical grinding, and EP (electropolishing) acid soaking for 24 hours, subjected to the hydrogen loading processing techniques, respectively.

Our findings indicate that increasing free hydrogen concentration in the cavity walls by approximately a factor of 2.5 results in a nearly tenfold reduction in the low-field quality factor, likely due to an increased volume fraction of niobium hydrides. These normal conducting inclusions inherit a diminished level of superconductivity via proximity-coupling to the niobium matrix and remain superconducting up to a breakdown field. Above this breakdown field, the hydrides become normal conducting and introduce significant loss, which increases with the field, a phenomenon known as the high-field Q slope (HFQS) [22]. We observe that significant hydrogen loading reduces the onset of HFQS from 25 MV/m to 0.4 MV/m. As the breakdown field inversely correlates with normal-metal inclusion size, distribution, and frequency (density), this lower onset field suggests significantly larger niobium hydride inclusions.

## III. DISCUSSION

Through the use of complementary characterization techniques, we revealed that nanometric-scale hydrides locally precipitate in the niobium pads of the superconducting qubit chip.

†zsung@fnal.gov

Topographical variations observed via cryo-AFM during cooling suggest that hydride precipitation in the Nb film follows the trend similar to those observed in bulk Nb [20,22,31]. The solid solubility of hydride becomes thermodynamically significant below 150 K [26,28] as confirmed by temperature-dependent grazing incident X-ray diffraction data presented in Fig. 3. Free hydrogen atoms located beneath the passive oxide layer are believed to be the primary resource for hydride phase formation, as indicated by ToF-SIMS depth profile.

Several possible scenarios exist for the causes of the hydrogen incorporation and subsequent hydride formation in the qubit Nb films. Firstly, hydrogen atoms may be incorporated during deposition due to residual hydrogen in the sputter chamber [9,35]. Similar to bulk Nb, any chemical treatment [36], annealing [37], or mechanical polishing [38,39] steps in the manufacturing process performed in a hydrogen-containing environment, which remove the protective $Nb_2O_5$ layer on the top of Nb, can lead to hydrogen incorporation in the underlying Nb. The recent study [25] highlights the critical role of the $Nb_2O_5$ layer on hydride formation in the Nb films, linked to degradation of the resonator for superconducting quantum qubit. In addition, a Cottrell atmosphere at dislocations or within dense dislocation networks may provide a favorable environment for hydride nucleation [40,41]. Therefore, further investigation is needed to identify the specific fabrication steps and post-fabrication structure defects contributing to hydrogen incorporation.

Given that ToF SIMS results confirm the presence of significant impurities such as oxygen, nitrogen, and carbon in the niobium—impurities which are known to act as effective hydrogen traps and thus suppress hydride formation [42–44]—the observed formation of Nb hydrides in certain regions of the qubit chip indicates that the current impurity concentrations may not yet be sufficient. Indeed, prior studies have demonstrated that increasing the oxygen concentration progressively reduces the H/Nb ratio and gradually eliminates hydride-related losses such as HFQS [45]. Therefore, strategically introducing higher concentrations of oxygen into the niobium films could further mitigate hydride formation, offering an effective pathway to enhance superconducting qubit performance.

A slight discrepancy in cryo-AFM topological features is observed between the Nb film and the bulk SRF Nb. As presented in Fig. 1, unlike in the bulk [31], distinct Nb hydride-related features did not appear during the initial cooldown but instead emerged after warming back from 2 K. Topographical features of ~ 2 nm in size were observed only during the initial cooldown below 50 K. This behavior suggests that the film's intrinsic columnar crystalline structure, as shown in the supplementary, may hinder hydrogen diffusion into defect sites within the Nb matrix, thereby suppressing hydride formation. Interestingly, our cryo-GIXRD indicates that the hydride morphology can evolve with repeated cooling cycles, consistent with findings in the previous study [20]. This observation may offer a possible explanation for changes in the qubit coherence times following each cooldown and the effect of "aging" from the accumulated number of cooldowns.

In superconducting RF Nb resonators, hydrogen incorporation within interstitial sites leads to the so-called "Q-slope" where the NbH phase, proximity coupled to the surrounding Nb matrix, causes significant degradation of the quality factor at high RF fields [21,22]. Furthermore, because these hydride regions exhibit a smaller superconducting gap (or $\Delta$, order parameter) than the surrounding Nb, they may act as sinks for quasiparticles and, in turn, diminish energy relaxation coherence times near the operating temperature. The widespread presence of these precipitates throughout the film explains that simply modifying film thickness is unlikely to mitigate this decoherence mechanism.

Fig. 4 describes that significant $Q_0$ degradation can also arise even in a low accelerating field (< 1 MV/m) at a lower temperature (< 1.6 K), similar to the quantum regime, as the precipitates of nanoscale hydride precipitation are coupled with the proximity effect of the superconducting to normal state. Of course, the extent of suppression of the quality factor depends on the amount of H-loading. In this study, the application of a 3D bulk Nb cavity makes it easy to understand the effects of hydride formation on the level of RF dissipation, compared to the study for the complex architecture of a superconducting quantum device. Surface property variation by engineering chemical and thermal process [46,47] has been sufficiently studied in the bulk Nb SRF cavity for optimal RF performance, which can be simply used for leveraging the complex structure device. This suggests that the substantially reduced decoherence time observed in the superconducting transmon qubit could be linked to the emergence of normal conducting hydride regions. However, our bulk Nb study does not address the potential involvement of two-level system (TLS) associated with the hydride phase. Further work should include detailed RF characterization using Nb film-based resonators.

In summary, our findings suggest that the Nb hydride precipitates in Nb thin films are pervasive and present a previously unaccounted-for source of the decoherence in superconducting qubits and planar Nb resonators. Niobium hydrides likely contribute to the form of quasiparticle dissipation and may contribute to cooldown-to-cooldown variability and long-term "aging" effects.

## ACKNOWLEDGMENTS

We thank Rigetti Computing and, in particular, Cameron Kopas and the Rigetti chip design and fabrication teams for the development and manufacturing of the qubit devices used in the reported experimental study. We acknowledge the Spanish Ministerio de Ciencia, Innovación y Universidades and Consejo Superior de Investigaciones Científicas for financial support through the projects 2010 6 0E 013 and 2021 60 E 030, and for the provision of synchrotron radiation facilities at BM25-SpLine at the ESRF. We are also grateful to Carlos Gerardo Torres Castanedo, Drs. Paul Chandrica Masih Das, Dominic Pascal Goronzy, Prof. Mark Hersam and David N Seidman. We would also like to thank Dr. Germán R. Castro and all the technical staff from the SpLine beam station for their valuable support during our experiments. This material is based upon work supported by the U.S. Department of Energy, Office of Science, National Quantum Information Science Research Centers, Superconducting Quantum Materials and Systems Center (SQMS) under contract number DE-AC02-07CH11359. This work made use of the EPIC, Keck-II, and/or SPID facilities of Northwestern University's NU*ANCE* Center, which received support from the Soft and Hybrid Nanotechnology Experimental (SHyNE) Resource (NSF ECCS-1542205); the MRSEC program (NSF DMR-1121262) at the Materials Research Center; the International Institute for Nanotechnology (IIN); the Keck Foundation; and the State of Illinois, through the IIN.

[†]zsung@fnal.gov

[1] M. A. Tanatar et al., *Anisotropic Superconductivity of Niobium Based on Its Response to Non-Magnetic Disorder*, arXiv:2207.14395.
[2] K. R. Joshi et al., *Quasiparticle Spectroscopy, Transport, and Magnetic Properties of Nb Films Used in Superconducting Transmon Qubits*, arXiv:2207.11616.
[3] A. Premkumar et al., Microscopic relaxation channels in materials for superconducting qubits, Commun. Mater. **2**, 72 (2021).
[4] C. Müller, J. H. Cole, and J. Lisenfeld, Towards understanding two-level-systems in amorphous solids: insights from quantum circuits, Rep. Prog. Phys. **82**, 124501 (2019).
[5] J. Zmuidzinas, Superconducting Microresonators: Physics and Applications, Annu. Rev. Condens. Matter Phys. **3**, 169 (2012).
[6] J. M. Martinis et al., Decoherence in Josephson Qubits from Dielectric Loss, Phys. Rev. Lett. **95**, 210503 (2005).
[7] R. McDermott, Materials Origins of Decoherence in Superconducting Qubits, IEEE Trans. Appl. Supercond. **19**, 2 (2009).
[8] D. P. Pappas, M. R. Vissers, D. S. Wisbey, J. S. Kline, and J. Gao, Two Level System Loss in Superconducting Microwave Resonators, IEEE Trans. Appl. Supercond. **21**, 871 (2011).
[9] A. A. Murthy, J. Lee, C. Kopas, M. J. Reagor, A. P. McFadden, D. P. Pappas, M. Checchin, A. Grassellino, and A. Romanenko, TOF-SIMS analysis of decoherence sources in superconducting qubits, Appl. Phys. Lett. **120**, 044002 (2022).
[10] D. Bafia, A. Murthy, A. Grassellino, and A. Romanenko, Oxygen vacancies in niobium pentoxide as a source of two-level system losses in superconducting niobium, Phys. Rev. Appl. **22**, 024035 (2024).
[11] M. V. P. Altoé et al., *Localization and Reduction of Superconducting Quantum Coherent Circuit Losses*, https://arxiv.org/abs/2012.07604v1.
[12] A. Romanenko and D. I. Schuster, Understanding Quality Factor Degradation in Superconducting Niobium Cavities at Low Microwave Field Amplitudes, Phys. Rev. Lett. **119**, 264801 (2017).
[13] A. Megrant et al., Planar superconducting resonators with internal quality factors above one million, Appl. Phys. Lett. **100**, 113510 (2012).
[14] C. T. Earnest, J. H. Béjanin, T. G. McConkey, E. A. Peters, A. Korinek, H. Yuan, and M. Mariantoni, Substrate surface engineering for high-quality silicon/aluminum superconducting resonators, Supercond. Sci. Technol. **31**, 125013 (2018).
[15] J. Wenner et al., Surface loss simulations of superconducting coplanar waveguide resonators, Appl. Phys. Lett. **99**, 113513 (2011).
[16] A. Romanenko, R. Pilipenko, S. Zorzetti, D. Frolov, M. Awida, S. Belomestnykh, S. Posen, and A. Grassellino, Three-Dimensional Superconducting Resonators at T < 20 mK with Photon Lifetimes up to $\tau$ = 2 s, Phys. Rev. Appl. **13**, 034032 (2020).
[17] K. Serniak, M. Hays, G. de Lange, S. Diamond, S. Shankar, L. D. Burkhart, L. Frunzio, M. Houzet, and M. H. Devoret, Hot Nonequilibrium Quasiparticles in Transmon Qubits, Phys. Rev. Lett. **121**, 157701 (2018).
[18] A. P. Vepsäläinen et al., Impact of ionizing radiation on superconducting qubit coherence, Nature **584**, 551 (2020).
[19] C. D. Wilen et al., Correlated charge noise and relaxation errors in superconducting qubits, Nature **594**, 369 (2021).
[20] F. Barkov, A. Romanenko, Y. Trenikhina, and A. Grassellino, Precipitation of hydrides in high purity niobium after different treatments, J. Appl. Phys. **114**, 164904 (2013).
[21] F. Barkov, A. Romanenko, and A. Grassellino, Direct observation of hydrides formation in cavity-grade niobium, Phys. Rev. Spec. Top. - Accel. Beams **15**, 122001 (2012).
[22] A. Romanenko, F. Barkov, L. D. Cooley, and A. Grassellino, Proximity breakdown of hydrides in superconducting niobium cavities, Supercond. Sci. Technol. **26**, 035003 (2013).
[23] Y. Trenikhina, A. Romanenko, J. Kwon, J.-M. Zuo, and J. F. Zasadzinski, Nanostructural features degrading the performance of superconducting radio frequency niobium cavities revealed by transmission electron microscopy and electron energy loss spectroscopy, J. Appl. Phys. **117**, 154507 (2015).
[24] T. Schober and H. Wenzl, *The Systems NbH(D), TaH(D), VH(D) : Structures, Phase Diagrams, Morphologies, Methods of Preparation*, in *Hydrogen in Metals II*, edited by G. Alefeld and J. Völkl, Vol. 29 (Springer Berlin Heidelberg, Berlin, Heidelberg, 1978), pp. 11–71.
[25] C. G. Torres-Castanedo et al., Formation and Microwave Losses of Hydrides in Superconducting Niobium Thin Films Resulting

†zsung@fnal.gov

[26] J. Hauck, Ordering of hydrogen in niobium hydride phases, Acta Crystallogr. Sect. A **33**, 208 (1977).
[27] B. Makenas and H. K. Birnbaum, On the techniques for preparing refractory metal — Hydrogen alloys for T.E.M. studies, Scr. Metall. **11**, 699 (1977).
[28] S. Isagawa, Hydrogen absorption and its effect on low-temperature electric properties of niobium, J. Appl. Phys. **51**, 4460 (1980).
[29] J. Knobloch, *The "Q Disease" in Superconducting Niobium RF Cavities*, in *AIP Conference Proceedings*, Vol. 671 (AIP, Newport News, Virginia (USA), 2003), pp. 133–150.
[30] A. Nersisyan et al., *Manufacturing Low Dissipation Superconducting Quantum Processors*, in *2019 IEEE International Electron Devices Meeting (IEDM)* (2019), p. 31.1.1-31.1.4.
[31] Z. Sung, A. Cano, A. Murthy, D. Bafia, E. Karapetrova, M. Martinello, J. Lee, A. Grassellino, and A. Romanenko, Direct observation of nanometer size hydride precipitations in superconducting niobium, Sci. Rep. **14**, 26916 (2024).
[32] J. Riffaud, M.-C. Lépy, Y. Ménesguen, and A. Novikova, Measurement of K fluorescence yields of niobium and rhodium using monochromatic radiation, X-Ray Spectrom. **46**, 341 (2017).
[33] L. G. Parratt, Surface Studies of Solids by Total Reflection of X-Rays, Phys. Rev. **95**, 359 (1954).
[34] O. Melnychuk, A. Grassellino, and A. Romanenko, Error analysis for intrinsic quality factor measurement in superconducting radio frequency resonators, Rev. Sci. Instrum. **85**, 124705 (2014).
[35] H. H. Shen, X. T. Zu, B. Chen, C. Q. Huang, and K. Sun, Direct observation of hydrogenation and dehydrogenation of a zirconium alloy, J. Alloys Compd. **659**, 23 (2016).
[36] T. Higuchi, K. Saito, Y. Yamazaki, T. Ikeda and S. Ohgushi, HYDROGENQ-DISEASEANDELECTROPOLISHING, The 10th Workshop on RF Superconductivity, 2001, Tsukuba, Japan **pr02**, (n.d.).
[37] K. Faber and H. Schultz, Hydrogen contamination in tantalum and niobium following UHV-degassing, Scr. Metall. **6**, 1065 (1972).
[38] T. Higuchi, G. U. A. Studies, and N. P. C. Ltd, Centrifugal Barrel Polishing of L-band Niobium cavities, 2 (2001).
[39] S. Balachandran, A. Polyanskii, S. Chetri, P. Dhakal, Y.-F. Su, Z.-H. Sung, and P. J. Lee, Direct evidence of microstructure dependence of magnetic flux trapping in niobium, Sci. Rep. **11**, 5364 (2021).
[40] A. H. Cottrell and B. A. Bilby, Dislocation Theory of Yielding and Strain Ageing of Iron, Proc. Phys. Soc. Sect. A **62**, 49 (1949).
[41] J. A. Rodrigues and R. Kirchheim, More evidence for the formation of a dense cottrell cloud of hydrogen (hydride) at dislocations in niobium and palladium, Scr. Metall. **17**, 159 (1983).
[42] B. Visentin, M. F. Barthe, V. Moineau, and P. Desgardin, Involvement of hydrogen-vacancy complexes in the baking effect of niobium cavities, Phys. Rev. Spec. Top. - Accel. Beams **13**, 052002 (2010).
[43] G. Pfeiffer and H. Wipf, The trapping of hydrogen in niobium by nitrogen interstitials, J. Phys. F Met. Phys. **6**, 167 (1976).
[44] B. Cox, Hydrogen trapping by oxygen and dislocations in zirconium alloys, J. Alloys Compd. **256**, L4 (1997).
[45] D. Bafia, A. Grassellino, and A. Romanenko, *The Role of Oxygen Concentration in Enabling High Gradients in Niobium SRF Cavities*, in *The Role of Oxygen Concentration in Enabling High Gradients in Niobium SRF Cavities* (US DOE, 2021).
[46] H. Padamsee, The science and technology of superconducting cavities for accelerators, Supercond. Sci. Technol. **14**, R28 (2001).
[47] Hasan Padamsee, Jens Knobloch, Tomas Hays, *RF Superconductivity for Accelerators, 2nd Edition | Wiley* (2008).

[25] ...from Fluoride Chemical Processing, Adv. Funct. Mater. **34**, 2401365 (2024).

†zsung@fnal.gov